\documentclass[conference]{IEEEtran}
\usepackage{times}

\usepackage[numbers]{natbib}
\usepackage{multicol}
\usepackage[bookmarks=true]{hyperref}
\usepackage{graphicx}

\begin{document}

\title{EMG-Driven Stiffness-Modulating Palpation for Telerehabilitation}


\author{\authorblockN{Thomas M. Kwok\authorrefmark{1},
Hilary HY Cheng\authorrefmark{2}, and
Wai Tuck Chow\authorrefmark{3}}
\authorblockA{\authorrefmark{1}Department of Mechanical and Mechatronics Engineering, 
University of Waterloo,
Canada\\ Email: thomasm.kwok@uwaterloo.ca}
\authorblockA{\authorrefmark{2}Department of Biomedical Engineering,
        National University of Singapore, Singapore\\
Email: hiuyeehilarycheng@gmail.com}
\authorblockA{\authorrefmark{3}School of Mechanical and Aerospace Engineering, Nanyang Technological University, Singapore\\
Email: wtchow@ntu.edu.sg}}

\maketitle

\IEEEpeerreviewmaketitle

\section{Introduction}
Telerehabilitation, the remote delivery of therapy using telecommunication technologies, has seen rapid growth, especially after the COVID-19 pandemic. It enables access to care for patients who are quarantined, immobile, or living far from clinical facilities \cite{munoz2023telerehabilitation,moulaei2022telerehabilitation,telerehab_covid}.

Although many robotic systems support remote rehabilitation \cite{Lanini2015,baur2019beam}, a major limitation is the lack of digital palpation. This hands-on method is essential for physiotherapists to assess muscle activation and stiffness \cite{davidson2020time}. Without it, therapists may find it difficult to evaluate muscle engagement remotely, and deliver a meaningful assistance for patients.

Kinesthetic feedback is essential for replicating palpation. Unlike cutaneous feedback, such as vibration or electrical stimulation, which provides indirect cues \cite{Prattichizzo2011CutaneousFF}, kinesthetic devices can simulate realistic touch by applying direct force to the finger \cite{see2022touch}. In telerehabilitation, these devices can render palpation by using force measurements when the remote robot presses on the patient’s muscle. However, latency in teleoperation causes desynchronization between motion commands and force feedback \cite{fu2018framework}, resulting in distorted and unreliable haptic sensations. Moreover, accurately assessing small or deep muscles is challenging when using a relatively large robot finger, further limiting the effectiveness of this approach.

Electromyography (EMG) sensors offer a promising alternative by measuring muscle engagement and activation, which reflects muscle stiffness and abnormal synergies \cite{emg_muscleStiff_Xie2020}. EMG is particularly useful for assessing small or deep muscles. However, the signal is typically displayed visually on a computer screen, which lacks tactile realism and may divide the therapist’s attention during assessment.

In previous work \cite{kwok_wearable_2025}, we developed a lightweight, wearable haptic device that uses a honeycomb jamming mechanism for stiffness rendering in teleoperated object-grasping tasks. In this abstract, we explore adapting this device into a palpation tool (HJ-Pal) for telerehabilitation. HJ-Pal modulates its stiffness to provide kinesthetic feedback based on EMG signals, allowing therapists to perceive muscle engagement remotely. Given the inherent noise in EMG signals, directly mapping these signals to device stiffness poses a challenge. As an initial step, we evaluate HJ-Pal’s capability to track EMG signals from small muscles, demonstrating its potential for remote muscle assessment. In future work, we plan to investigate the correlation between EMG activity and the device’s rendered stiffness, validating its effectiveness for digital palpation in remote rehabilitation contexts.

\section{Relevance to HRCM}

This work contributes to the goals of the Workshop on Human-Robot Contact and Manipulation (HRCM 2025) by presenting a novel application of wearable haptics for physical human-robot interaction (pHRI) in remote healthcare. HJ-Pal offers a new approach to remote palpation during telerehabilitation by using a honeycomb-jamming mechanism to modulate stiffness in response to EMG signals. This allows therapists to physically perceive muscle engagement, even in small or deep muscles, bridging the gap between traditional hands-on assessment and remote therapy. Unlike conventional visual EMG displays, HJ-Pal reintroduces tactile perception into the therapist’s workflow, enhancing clinical intuition and reducing cognitive load. By introducing HJ-Pal, we aim to open new directions in remote patient monitoring by using haptics that go beyond surgical applications.

To advance this research, we aim to investigate how kinesthetic feedback influences therapist decision-making, user trust, and perceived effectiveness in real-world clinical environments. This work serves as both a contribution and a starting point for discussions within the workshop on how pHRI frameworks can be extended to address the evolving demands of remote and home-based healthcare.

\section{HJ-Pal for Remote Palpation}

HJ-Pal is a lightweight, thumb-wearable haptic device that leverages a honeycomb jamming mechanism to deliver kinesthetic feedback for remote muscle assessment in telerehabilitation. It builds upon our previous work on stiffness-modulating haptics \cite{kwok_wearable_2025}. By adjusting vacuum pressure, the honeycomb structure modulates its stiffness in real time. For detailed mechanical design and characterization, refer to \cite{kwok_wearable_2025}.

Unlike traditional haptic systems that rely on delayed force feedback from remote robots, HJ-Pal modulates stiffness locally based on EMG signals, enabling therapists to perceive muscle activation through direct touch.

\begin{figure}[!t]
\centerline{\includegraphics[width=0.9\columnwidth]{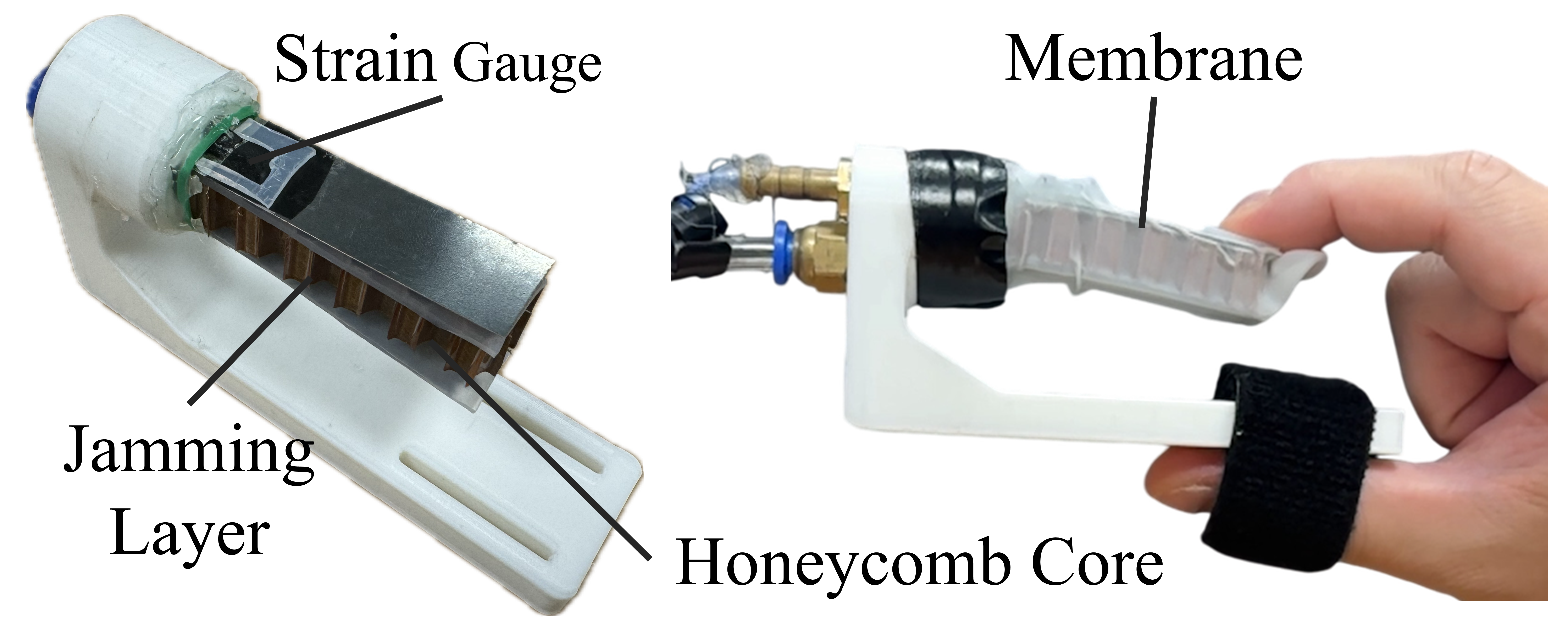}}
\caption{The prototype of honeycomb-jamming palpation device (HJ-Pal).}
\label{fig:hj_pal}
\end{figure}

\begin{figure}[!t]
\centerline{\includegraphics[width=1.0\columnwidth]{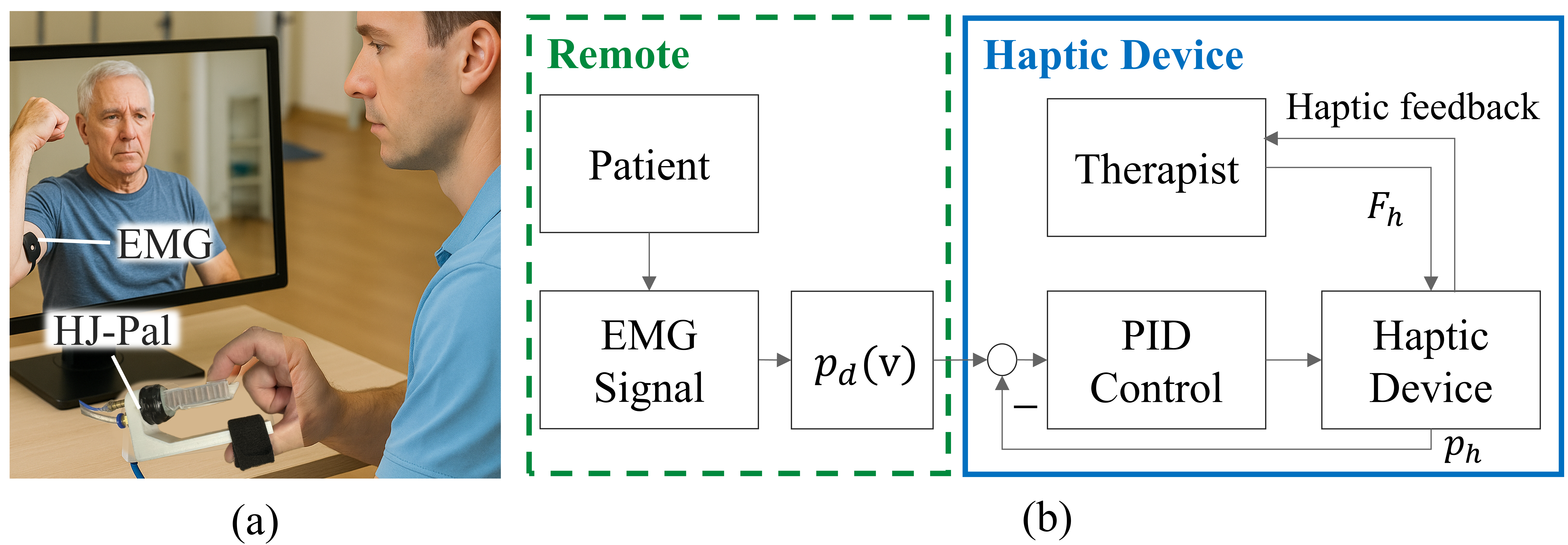}}
\caption{(a) Conceptual illustration of a telerehabilitation scenario using HJ-Pal, generated with Sora by OpenAI \cite{sora2024}. (b) The proposed remote palpation framework for HJ-Pal. Here, $p_d$ denotes the desired pressure, $p_h$ is the measured feedback pressure, and $F_h$ represents the fingertip force applied by the therapist.}
\label{fig:teleopt}
\end{figure}

As shown in Fig.\ref{fig:hj_pal}, the device incorporates a honeycomb core, membrane and jamming layers, enabling stiffness modulation up to 0.85N/mm under vacuum pressure. A strain gauge captures finger indentation, while a PID-controlled vacuum pump adjusts internal pressure and stiffness of honeycomb-jamming mechanism based on EMG signals. The overall remote palpation framework is illustrated in Fig.\ref{fig:teleopt}. When therapists press on the device, they experience kinesthetic haptic feedback that reflects the patient’s muscle activation. When released, HJ-Pal automatically returns to its original, flexible state.

As shown in Fig.~\ref{fig:teleopt}, we use EMG sensors (Delsys Trigno) to measure muscle activation and map the processed signal to the desired pressure ($p_d$) within HJ-Pal’s available stiffness range. The raw EMG data are filtered using a 4th-order Butterworth bandpass filter (10–500Hz), then detrended, rectified, and smoothed with a moving mean filter. The resulting signal is normalized to the participant’s Maximum Voluntary Contraction (MVC).

This is the novel application of EMG-driven jamming in a digital palpation framework. It can potentially integrate with upper limb exoskeleton \cite{Kwok2024, Kwok2024_2}. In this case, HJ-Pal enables therapists to sense patient muscle engagement and adjust robotic assistance, offering a interactive physical human-robot interaction for telerehabilitation.

\subsection{Results \& Discussion}

As an initial step toward developing HJ-Pal as a palpation tool for remote muscle assessment, we conducted a preliminary experiment to evaluate its ability to render EMG-driven stiffness modulation via pressure tracking. Given the difficulty in assessing small muscles through traditional palpation, we selected the pronator and supinator muscles as targets.

Two EMG sensors (placed according to SENIAM guidelines) recorded muscle activation from a single participant. During the experiment, the participant rested his forearms on a table in a neutral position and performed five cycles of slow forearm supination (rotating palms upward), briefly holding the position before returning to neutral and relaxing.

\begin{figure}[!t]
\centerline{\includegraphics[width=1.0\columnwidth]{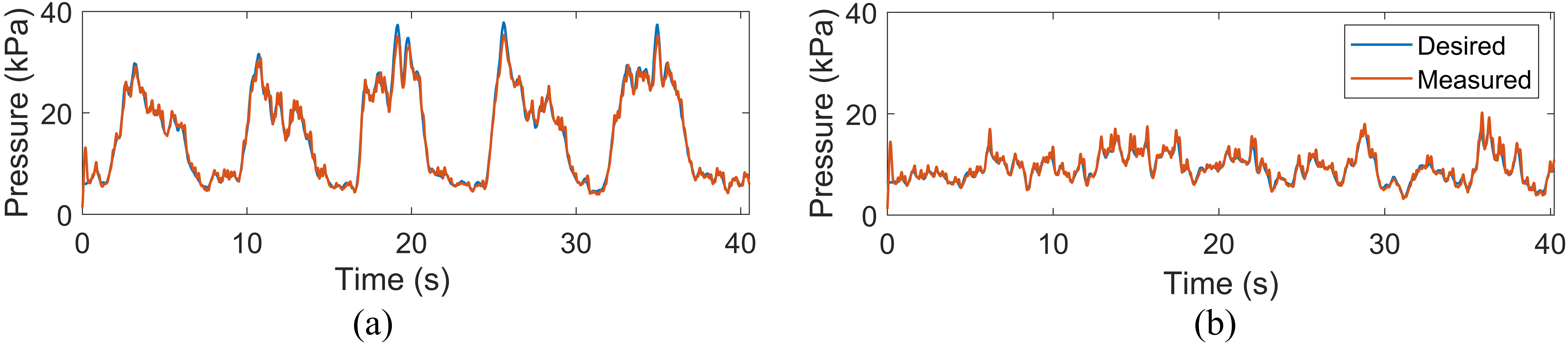}}
\caption{Vacuum pressure tracking of EMG-derived commands during forearm supination: (a) supinator and (b) pronator muscle activation.}
\label{fig:result}
\end{figure}

The processed EMG signals were mapped to desired pressure values ($p_d$), which were tracked by the PID-controlled vacuum system of HJ-Pal. As shown in Fig.~\ref{fig:result}, the measured internal pressure closely followed the EMG-derived pressure commands. The system achieved low root-mean-square error (RMSE) values of $1.006±0.031$kPa in supinator and $0.904±0.056$kPa in pronator. Furthermore, the measured pressure correlated strongly with the EMG-derived commands (Pearson’s correlation with $r=0.994±0.0004$ in supinator; $r=0.966±0.010$ in pronator, all $p<0.0001$), confirming that the HJ-Pal is capable of responding to the high-frequency, noisy nature of EMG input.

These results suggest that HJ-Pal can deliver reliable kinesthetic feedback based on muscle activity. Such fidelity is essential for enabling therapists to perceive subtle variations in muscle engagement—especially in small or deep muscles—through haptic feedback. This functionality positions HJ-Pal as a promising tool for remote muscle assessment in telerehabilitation settings, where traditional force feedback mechanisms may be delayed, distorted, or infeasible.

Future work will explore the correlation between EMG activity and HJ-Pal's perceived stiffness, further validating HJ-Pal’s application for digital palpation. We also plan to conduct user studies to assess the usability of HJ-Pal in realistic telerehabilitation scenarios. In particular, we aim to compare EMG-based haptic feedback with traditional approaches such as visualizing real-time EMG signals on a computer screen and employing force-based haptic feedback. Comparative metrics will include diagnostic accuracy in identifying muscle activation levels and the user's cognitive load. 

\section{Conclusion}
This work presents HJ-Pal, a lightweight, thumb-wearable haptic device that uses honeycomb jamming to render kinesthetic feedback based on EMG signals. Preliminary results show reliable pressure tracking and strong correlation with muscle activity, demonstrating HJ-Pal’s potential for remote palpation and assessment of small muscles in telerehabilitation.

\bibliographystyle{plainnat}
\bibliography{references}

\begin{thebibliography}{14}
\providecommand{\natexlab}[1]{#1}
\providecommand{\url}[1]{\texttt{#1}}
\expandafter\ifx\csname urlstyle\endcsname\relax
  \providecommand{\doi}[1]{doi: #1}\else
  \providecommand{\doi}{doi: \begingroup \urlstyle{rm}\Url}\fi

\bibitem[Aderonmu(2020)]{telerehab_covid}
Joseph~Ayotunde Aderonmu.
\newblock Emerging challenges in meeting physiotherapy needs during covid-19 through telerehabilitation.
\newblock \emph{Bulletin of Faculty of Physical Therapy}, 25:\penalty0 1--4, 2020.

\bibitem[Baur et~al.(2019)Baur, Rohrbach, Hermsd{\"o}rfer, Riener, and Klamroth-Marganska]{baur2019beam}
Kilian Baur, Nina Rohrbach, Joachim Hermsd{\"o}rfer, Robert Riener, and Verena Klamroth-Marganska.
\newblock The “beam-me-in strategy”--remote haptic therapist-patient interaction with two exoskeletons for stroke therapy.
\newblock \emph{Journal of neuroengineering and rehabilitation}, 16:\penalty0 1--15, 2019.

\bibitem[Davidson et~al.(2020)Davidson, Nielsen, Taberner, and Kruger]{davidson2020time}
Melissa~J Davidson, Poul~MF Nielsen, Andrew~J Taberner, and Jennifer~A Kruger.
\newblock Is it time to rethink using digital palpation for assessment of muscle stiffness?
\newblock \emph{Neurourology and urodynamics}, 39\penalty0 (1):\penalty0 279--285, 2020.

\bibitem[Fu et~al.(2018)Fu, van Paassen, Abbink, and Mulder]{fu2018framework}
Wei Fu, Marinus~M van Paassen, David~A Abbink, and Max Mulder.
\newblock Framework for human haptic perception with delayed force feedback.
\newblock \emph{IEEE Transactions on Human-Machine Systems}, 49\penalty0 (2):\penalty0 171--182, 2018.

\bibitem[Kwok and Yu(2024{\natexlab{a}})]{Kwok2024}
Thomas~M. Kwok and Haoyong Yu.
\newblock A novel bilateral underactuated upper limb exoskeleton for post-stroke bimanual adl training.
\newblock \emph{IEEE Transactions on Neural Systems and Rehabilitation Engineering}, 32:\penalty0 3299--3309, 2024{\natexlab{a}}.
\newblock \doi{10.1109/TNSRE.2024.3407653}.

\bibitem[Kwok and Yu(2024{\natexlab{b}})]{Kwok2024_2}
Thomas~M. Kwok and Haoyong Yu.
\newblock Asymmetric bimanual adl training with underactuated exoskeleton using independent joint control and visual guidance.
\newblock \emph{IEEE Access}, 12:\penalty0 9277--9291, 2024{\natexlab{b}}.
\newblock \doi{10.1109/ACCESS.2024.3352911}.

\bibitem[Kwok et~al.(2025)Kwok, Zhang, and Chow]{kwok_wearable_2025}
Thomas~M. Kwok, Bohan Zhang, and Wai~Tuck Chow.
\newblock A {Wearable} {Stiffness}-{Rendering} {Haptic} {Device} with a {Honeycomb} {Jamming} {Mechanism} for {Bilateral} {Teleoperation}.
\newblock \emph{Machines}, 13\penalty0 (1):\penalty0 27, January 2025.
\newblock ISSN 2075-1702.
\newblock \doi{10.3390/machines13010027}.

\bibitem[Lanini et~al.(2015)Lanini, Tsuji, Wolf, Riener, and Novak]{Lanini2015}
Jessica Lanini, Toshiaki Tsuji, Peter Wolf, Robert Riener, and Domen Novak.
\newblock Teleoperation of two six-degree-of-freedom arm rehabilitation exoskeletons.
\newblock In \emph{2015 IEEE International Conference on Rehabilitation Robotics (ICORR)}, pages 514--519, 2015.
\newblock \doi{10.1109/ICORR.2015.7281251}.

\bibitem[Moulaei et~al.(2022)Moulaei, Sheikhtaheri, Nezhad, Haghdoost, Gheysari, and Bahaadinbeigy]{moulaei2022telerehabilitation}
Khadijeh Moulaei, Abbas Sheikhtaheri, Mansour~Shahabi Nezhad, AliAkbar Haghdoost, Mohammad Gheysari, and Kambiz Bahaadinbeigy.
\newblock Telerehabilitation for upper limb disabilities: a scoping review on functions, outcomes, and evaluation methods.
\newblock \emph{Archives of Public Health}, 80\penalty0 (1):\penalty0 196, 2022.

\bibitem[Mu{\~n}oz-Tom{\'a}s et~al.(2023)Mu{\~n}oz-Tom{\'a}s, Burillo-Lafuente, Vicente-Parra, Sanz-Rubio, Suarez-Serrano, Marc{\'e}n-Rom{\'a}n, and Franco-Sierra]{munoz2023telerehabilitation}
M{\textordfeminine}~Teresa Mu{\~n}oz-Tom{\'a}s, Mario Burillo-Lafuente, Araceli Vicente-Parra, M{\textordfeminine}~Concepci{\'o}n Sanz-Rubio, Carmen Suarez-Serrano, Yolanda Marc{\'e}n-Rom{\'a}n, and M{\textordfeminine}~{\'A}ngeles Franco-Sierra.
\newblock Telerehabilitation as a therapeutic exercise tool versus face-to-face physiotherapy: a systematic review.
\newblock \emph{International Journal of Environmental Research and Public Health}, 20\penalty0 (5):\penalty0 4358, 2023.

\bibitem[OpenAI(2024)]{sora2024}
OpenAI.
\newblock Sora.
\newblock \url{https://openai.com/sora}, 2024.
\newblock Video and image generation platform.

\bibitem[Prattichizzo et~al.(2011)Prattichizzo, Pacchierotti, and Rosati]{Prattichizzo2011CutaneousFF}
Domenico Prattichizzo, Claudio Pacchierotti, and Giulio Rosati.
\newblock Cutaneous force feedback as a sensory subtraction technique in haptics.
\newblock \emph{IEEE Transactions on Haptics}, 5:\penalty0 289--300, 2011.

\bibitem[See et~al.(2022)See, Choco, and Chandramohan]{see2022touch}
Aaron~Raymond See, Jose Antonio~G Choco, and Kohila Chandramohan.
\newblock Touch, texture and haptic feedback: a review on how we feel the world around us.
\newblock \emph{Applied Sciences}, 12\penalty0 (9):\penalty0 4686, 2022.

\bibitem[Xie et~al.(2020)Xie, Leng, Zhi, Jiang, Tian, Luo, Yu, and Song]{emg_muscleStiff_Xie2020}
Tian Xie, Yan Leng, Yihua Zhi, Chao Jiang, Na~Tian, Zichong Luo, Hairong Yu, and Rong Song.
\newblock Increased muscle activity accompanying with decreased complexity as spasticity appears: High-density emg-based case studies on stroke patients.
\newblock \emph{Frontiers in Bioengineering and Biotechnology}, 8, 2020.
\newblock ISSN 2296-4185.
\newblock \doi{10.3389/fbioe.2020.589321}.

\end{thebibliography}

\end{document}